# The quantum motion as the motion in the integrable Weyl space.

## A.V.Rogachev.

Experimental Laboratory, JSC "Poltava Diamond Plant"
71A Krasina str., Poltava, Ukraine, 36023.

A new version of hidden variables theory founded on the generalization of world's geometry is proposed. The quantum-mechanical motion as the motion in some "inner space", which has a structure of the integrable Weyl space is examined. Equations of motion for quantum particles as well as equations for the "quantum-mechanical field", "guiding" the particle, are deduced. The wave equation for ensembles, the Born's interpretation of wave function and the "guiding equation" as the consequences of proposed model are obtained.

03.65.Ud;   03.65.Ta.

## 1. Introduction.

The theme of this article is the Bohm (or de Broglie-Bohm) theory [1,2] - the oldest and surely the most suitable from so-called "theories of hidden variables" (THV) [3]. In this theory – like the classical and unlike the usual quantum mechanics – the quantum particle has in every moment of time the exactly defined position and velocity and its motion is chorerographed in deterministic manner by the some additional "quantum-mechanical field" $\Psi$ .

The Bohmian mechanics accounts for all of phenomena of nonrelativistic quantum mechanics, and besides makes this in many cases in more transparent form as the ortodoxal kopenhagen interpretation with its traditional invocation of status of observation.

Nevertheless, in spite of all these attractive properties, the most of physicists until now continues to consider the Bohmiam mechanics rather as the somewhat trick as the "good" physical theory. The main cause, why it happens so, is, in our opinion, the strongly marked phenomenological character of this theory. In particular, the equation of motion for particles (i.e. the Hamilton-Jacoby equation with the "quantum potential" in the original Bohm's paper [2,4,5] or the "guiding equation" in the some equivalent formulations of Bohm theory [6,7]) as well as the equation of Schroedinger, which rules the dynamics of "quantum-mechanical field", are not derived here from the some fundamental principle, but postulated in such manner, to have a full agreement with the quantum mechanics. The same can be said also about the connection between the behavior of quantum ensembles and the dynamics of individual systems. The Born's rule $w \propto |\Psi|^2$, which connects in Bohm theory the "quantum-mechanical field" $\Psi$ of individual system and the probability's density $w$ of ensemble, to which this system belongs, requires for its derivation the additional "hypothesis of quantum equilibrium" [6,8,9], justification of which remains until now rather delicate matter.

In this article the not complicated geometrical construction is offered, within the "rather strange and arbitrary" [10] "quantum potential" as well as the Schroedinger equation[1] and the Born's interpretation of $\Psi$ emerge in the most natural way with no using of any artificial assumptions. But for this we'll have to extend our knowing about the physical space and pass from the classical Riemannian space, properties of which are the same for all particles, to "inner space", which has the structure of integrable Weyl space (IW-space) [11] and properties of which are determined for every particle individually.

---

[1] more exactly - its relativistic analog, i.e. the Klein-Fock's equation

The model, offered here can't be called the exhaustive geometrical interpretation of Bohmian mechanics and is limited in few aspects.
1. First of all it is "semi-classical" i.e. the quantum features are assigned only to the particles but not to the fields.
2. It describes only the relatively weak quantum effects and does not account for phenomenon such as particle's creation and annihilation.
3. It considers only the case of particle with the spin zero.
4. The model is only the one-particle theory and thus leaves without consideration most of quantum non-local effects.

Regarding points 1 and 2, its successive calculation is rather the task for the Bohmian field theory [4,5,12], as for the Bohmian mechanics. The points 3 and 4 we prefer to discus somewhere else, since for its account the considerable and in this stage undesirable complication of primary scheme is required. [2]

## 2. Inner space.

The geometrical background of our model is the Weyl geometry, which admits the changing of vector's length by the parallel displacement. The main notions and relations of Weyl geometry are presented in Appendix A. Details may be found for instance in [15-17]. The Appendix A is necessary for understanding of this article, since we will use the Weyl's formalism very actively.

If in the Riemannian geometry a length of vector $x^i$ is not changed by the parallel displacement from the point $x^i$ to the point $x^i + dx^i$, i.e.
$$d(g^{ik}x_i x_k) = 0 \tag{2.1}$$
then in the Weyl geometry we have
$$d(g^{ik}x_i x_k) = 2(g^{ik}x_i x_k)k_l dx^l \tag{2.2}$$
where $k^l$ is so called scale vector, describing together with $g_{ik}$ the metrical properties of the Weyl space. Applying (2.2) to the square of interval $ds^2 = g_{ik}dx^i dx^k$ we find out, that in the Weyl geometry the length of scale must change by the parallel displacement, and hence it is impossible here (differing from the Riemannian space's case) to enter the "universal scale" for the whole space-time by means of simple translation of the "standard". Instead of this one should introduce in all of points of space-time its own "standard" for measuring of intervals, that in consequence of indefiniteness of such procedure leads to the requirement of scale invariance for every physical theory, based on the Weyl geometry. In other words, the dynamical equation of such theory must be invariant under the scale transformations
$$ds \to d\overline{s} = l(x)ds \iff g_{ik}(x) \to \overline{g}_{ik}(x) = l^2(x)g_{ik}(x) \tag{2.3}$$
(where $l(x)$ is an arbitrary scalar function), which are the mathematical equivalent of the passage from the one standard of scale to the another one.[3]

---
[2] It should be mentioned, that the task of construction of many-particle relativistic Bohmian mechanics remains until now unsolved. Some approach in this direction can be found in [ 4,9,13,14 ].

[3] As was been mentioned by Dirac [15] and also by Canuto, Adams et al. [11], the scale invariance must be (together with the coordinate one) the basis for *every* physical theory, since the choice of standard is not given by the nature, but depends only upon us. In this article we accept this Dirac's idea and will construct our model at once in the scale-invariant form.

One of the main moments of the Weyl's theory of 1918 [18,19] is an identification of scale vector $k_m$ with a vector potential of electromagnetic field and due to this - as Weyl and Dirac showed [18,19,15] - an unified description of gravitation and electromagnetism is possible. It is well known although [20] that when using an ordinary procedure of measurement of intervals with help of hard rules and usual clocks the physical space behaves itself as integrable i.e. two scales coinciding in the point $A$ will coincide in any other point $B$, even if they were transported in $B$ along different ways. If to assume that the geometry of world must reflect a behavior of real scales then to properties of vector $k_m$ a condition of integrability must be given.

$$k_{l;m} - k_{m;l} = 0 \tag{2.4}$$

from which follows, that the scale vector $k_m$ must be a gradient of a scalar function and hence it can not be identified with a potential of electromagnetic field. Unlike Weyl and Dirac, we will connect $k_m$ not with electromagnetic field, but with a "quantum-mechanical" one, which, as it will be shown further, doesn't require a declining of condition of integrability.

Another possible interpretation of $k_m$, differing both from Weyl's one and ours one war offered by authors of "the scale-covariant theory of gravitation" (SCTG) [11]. The SCTG not only rejects the electromagnetic nature of scale vector and admits the integrability of space, but goes further and declares the scale vector as a gauge degree of freedom, which describes only the difference between used system of units and the "privileged" one, in which not only the integrability takes place, but the keeping of the scale under parallel displacement as well.

Differing from this "gauge interpretation" of $k_m$ we will proceed from the assumption about physical nature of scale vector, that on the one hand will mean a presence of dynamical equations for it, and on the other hand, a direct influence of $k_m$ on the dynamics of particles. Namely, we'll postulate the **Assumption A**, that *for every particle there exists some individual "inner" IW- space, properties of which are described by the metric tensor $g_{lm}$, which is common for all particles and equal to the metric tensor of the classical space, and by the scale vector $k_m$, which is an individual characteristic of particle and is defined for every particle separately.*

It should be mentioned, that the **Assumption A,** that supposes the equality of metric tensors of classical and inner spaces is not logically necessary. One can show e.g., that the alternative **Assumption B**

$$k_m(inner) = k_m(klassical), \quad g_{mn}(inner) = \boldsymbol{b}^2(x) g_{mn}(klassical)$$

(where $\boldsymbol{b}(x)$ is some scalar function) leads in consequence to the same equation as the **Assumption A**. However, the **Assumption B** is absolutely unsatisfactory from the methodological considerations. Just as the $g_{lm}(klassical)$ can be always measured immediately, there exists no method of direct measuring of $g_{mn}(inner)$. More precisely: the classical part of $g_{mn}(inner)$ (i.e. $g_{lm}(klassical)$) can be measured by means of classical rulers and clocks, and the factor $\boldsymbol{b}^2(x)$ - only from the observation of quantum peculiarities of motion. Such a mixture of quantum and classical aspects in one and the same physical value looks very unnatural – unlike from the interpretation, founded on the **Assumption A**, where $g_{lm}$ is simply the classical gravitational field and $k_m$ - the new "inner" degree of freedom answering for the quantum phenomenon.

# 3. The motion of test particle.

In the classical mechanics, which assumes, that all particles move in a same "external space" with the same $g_{lm}$ and $k_m$, the scale invariant action for a test particle looks as follows (see for instance [15])

$$S = -\int m\boldsymbol{b}(x)ds - e\int A_i dx^i \tag{3.1}$$

where $e$ and $m$ are constants[4], $A_m$ is the potential of electromagnetic field and $\boldsymbol{b}$ is a some scalar function. This function is common for all particles and should be identified with the scale factor (see (A16)) of classical IW-spaces in order to equations of motion in absence of electromagnetic field coincide with the geodesic equation in IW-space [11]. In the expression (3.1) the values $m$, $e$ and $A_m$ are invariant under scale transformations, and the scale factor $\boldsymbol{b}$ is a co-scalar of power $-1$, that in total gives a scale invariance of (3.1). The "privileged" system of units $\boldsymbol{b} \equiv 1$, in which the length of vector doesn't change by the parallel displacement, coincides with the usual atomic system in which the masses of particles are constant.

Further, as it was mentioned above, we will try to explain quantum phenomenon, as a result of motion in the inner space, and for this we'll have to pass from action (3.1) to the action of more general form. To this "quantum" action will be given the natural requirements as follow
1. It is invariant under coordinate und scale transformations;
2. The scale vector of inner space must be included into quantum action explicitly;
3. In the classical limit $\hbar \to 0$ the quantum action must pass into the action (3.1) for a classical particle.

To all of these requirements the action as follows satisfies

$$S = -\int (m^2 \boldsymbol{b}^2 + \boldsymbol{a}R)^{1/2} ds - e\int A_i dx^i \tag{3.2}$$

where $m$, $e$, $A_m$ and $\boldsymbol{b}$ have the same meaning as before, $R$ is a scalar curvature of inner space (see (A.7)) and $\boldsymbol{a}$ is a constant, characterizing a degree of connection with "inner curvature"[5].

In case if $\boldsymbol{a} \to 0$, the action (3.2) is transformed into the ordinary classical action for a particle with a charge $e$ and a mass $m\boldsymbol{b}$.

The variation of action (3.2) in coordinates of particle gives

$$dS = -\int_a^b \{\boldsymbol{m}_{,l} + \frac{1}{2}\boldsymbol{m}g_{ik,l}u^i u^k - (\boldsymbol{m}g_{il}u^i)_{,k}u^k + eF_{li}u^i\}dx^l ds - (\boldsymbol{m}u_l + eA_l)dx^l \Big|_a^b \tag{3.3}$$

where $u^l$ is a 4-velocity of particle, $F_{li} = A_{i,l} - A_{l,i}$ is the tensor of electromagnetic field and $\boldsymbol{m}$ is a "quantum mass" of particle, which is given as follows

$$\boldsymbol{m}^2 = m^2 \boldsymbol{b}^2 + \boldsymbol{a}R \tag{3.4}$$

Supposing, that the variations in coordinates on limits of integration are equal to zero we can obtain the equations of motion as follows

---

[4] In this article we'll call a mass not a constant $m$, but the production $m\boldsymbol{b}$. So the mass is a co-scalar of power $-1$.

[5] A connection of constant $\boldsymbol{a}$ with the Plank's constant will be found in the section 7.

$$u_{l;m}u^m = \frac{\boldsymbol{m}_{,m}}{\boldsymbol{m}}(\boldsymbol{d}_l^m - u_l u^m) + \frac{e}{\boldsymbol{m}}F_{lm}u^m \tag{3.5}$$

(where ";" is an ordinary derivative on Christoffel's symbols) or in terms of co-tensors (see (A 11-12):

$$u_{l*m}u^m = \frac{\boldsymbol{m}_{*m}}{\boldsymbol{m}}(\boldsymbol{d}_l^m - u_l u^m) + \frac{e}{\boldsymbol{m}}F_{lm}u^m \tag{3.6}$$

We state that the equations (3.6) are the equations of motion of quantum particle in external gravitational and electromagnetic fields. The scale vector $k_m$ of inner space and the constant $\boldsymbol{a}$ are included in these equations through the "quantum mass" $\boldsymbol{m}$, that makes a particle to diverge from a classical trajectory.

Passing into (3.5) to the classical limit $\boldsymbol{a} \to 0$ gives the equations

$$u_{l;m}u^m = \frac{\boldsymbol{b}_{,m}}{\boldsymbol{b}}(\boldsymbol{d}_l^m - u^m u_l) + \frac{e}{m\boldsymbol{b}}F_{lm}u^m \tag{3.7}$$

which don't contain the scale vector of inner space anymore, and in absence of electromagnetic field coincide with the geodesic equations in the classical IW-space [11].

Further we need also the equation of Hamilton-Jakoby, which can be obtained, if to put $(\boldsymbol{dx})_a = 0, (\boldsymbol{dx})_b = \boldsymbol{dx} \neq 0$ in (3.3) and to suppose, that the passage from $a$ to $b$ takes place only along the "true" trajectory, which satisfies to the equations (3.6). In this case we find

$$\boldsymbol{dS} = -(\boldsymbol{m}u_l + eA_l)\boldsymbol{dx}^l$$

that leads to the Hamilton-Jacoby's equation

$$g^{lm}(S_{,l} + eA_l)(S_{,m} + eA_m) = \boldsymbol{m}^2 = m^2\boldsymbol{b}^2 + \boldsymbol{a}R \tag{3.8}$$

in which the scale vector of inner space is included in the scalar curvature $R$. In the classical limit $\boldsymbol{a} \to 0$ the equation (3.8) does not involve the scale vector $k_m$ anymore and precisely coincides with the classical Hamilton-Jacoby equation.

### 4. The equation for the $k$-field.

To calculate the trajectory of the quantum test particle it is not sufficient to know only the equations of motion (3.6) and the classical fields $A^i$ and $g_{ik}$, acting on the particle; it is necessary to know also the metric vector $k_i$ of particle's inner space, that appears in (3.6).

Since the metric vector $k_i$ is considered in our model as a new physical field but not as the some additional gauge degree of freedom, so its values $k_i(x,t)$ which are substituted in (3.6) can not be already arbitrary (as in the SCTG [11]), but should satisfy some "field equations", which in its turn should be deduced by means of variation with respect to $k_i$ from some "field action".

To avoid misunderstanding it should be noted, that the further reasoning is not a deducing of such additional field action. The latter can't be deduced, it should be postulated! But the further reasoning at least will help us understand why one should postulate this action exactly as we have done it and not in any other manner.

At first, let's note an evident fact, namely, that the action (3.2) describing the motion of the test particle is absolutely inappropriate for deducing of any field equations. In fact, every field value in the field equations should be defined in the whole 4-dimentional space-time. At the same time all values in the action (3.2) are defined only along a single world line. Hence, follows **Requirement 1:**

*The action I for the k-field should be written down as an integral over the whole 4-dimentional space-time.*

So simple is also **Requirement 2:**

*The action I for the k-field should disappear in absence of the particle the inner space of which is described by it*

Or, in other words, for the $k_i$ there should not exist any "equations of free field", which exist for any "normal" physical fields. The origin of Requirement 2 is quite evident: in fact, it is fully nonsensically to speak about the "inner space" of particle in absence of particle itself.

In order to formulate the **Requirement 3** let's assume, that the metric vector $k_i$ and the 4-velocity $u^l$ of the particle should be connected somehow with the "quantum-mechanical field" $\Psi_B$ of Bohmian mechanics. This assumption looks quite natural, since $k$ and $\Psi_B$ execute the same function of "guiding field" and the 4-velocity $u_i(x,t)$ is constrained in the BM by the $\Psi_B$ through the "guiding equation" [6,7]. This assumption can be written down symbolically as

$$\Psi_B(x,t) \Rightarrow k_i(x,t), \qquad (4.1)$$
$$\Psi_B(x,t) \Rightarrow u_i(x,t) \quad \text{or} \quad \Psi_B(x,t) \Rightarrow S(x,t),$$

where $S$ is a solution of Hamilton-Jacoby equation for given system.

For the quantum-mechanical field $\Psi_B$ in BM the following two properties are satisfied

1. Every measurement over the given system changes the $\Psi$ –field of system.

2. If two individual systems $M$ and $N$, each consisting of a test particle and of some "classical environment" (i.e. of some distribution of sources of classical fields $A^i$ and $g_{ik}$) are "identically prepared"[6] (i.e. for the test particles we have $e^{(N)} = e^{(M)}$, $(mb)^{(N)} = (mb)^{(M)}$ and for the classical environment - $A_i^{(N)} = A_i^{(M)}$, $g_{ik}^{(N)} = g_{ik}^{(M)}$), then for these two systems it should take place $\Psi^{(N)} = \Psi^{(M)}$, i.e. the equality of $\Psi$-fields of test particles.

If two groups of electrons with an absolute identical history (e.g., created in two identical electronic tubes) fall onto two identical double-splits, then the diffraction patterns on two identical screens located identically towards the splits will always be absolutely identical. But it is enough to change even slightly something in arrangement of one of experimental plants (for example, the accelerating potential of electronic tube), and the identity of diffraction pattern will be lost.

By force of (4.1) these properties can be extended immediately on $k_i$ and $S$-values. Thus, if we wish to define $k_i$ and $S$ for some system, then we should not make any measurements over this system itself, but only over another systems, which are identical to the given. However, if the observer declines some measurements over given system, then he loses any possibility to obtain any data about initial coordinates of particle, without which it is impossible to calculate the trajectory of particle. In other words, if one knows $k_i$ and $S$, then one can't know the initial coordinates and vice versa, if the initial coordinates $\vec{x}_0, t_0$ are certainly defined, then $k_i$ and $S$ become absolutely indefinite.

---

[6] The more precise definitions of two systems identity is given in the Appendix B.

Now, one can formulate

**The requirement 3**: *The action I for k-field should take into account our ignorance about the initial coordinates of particle*

i.e. it should be constructed in such a manner as if we knew what trajectories is possible for the particle with given $\Psi_b$, but don't know, along which of them the particle moves in reality.

All formulated above requirements 1-3 are satisfied for the action

$$I = \int d^3\mathbf{l}\, P(\mathbf{l}) \int_{L(\mathbf{l})} ds\{\mathbf{m} + eA_i u^i\} \tag{4.2}$$

which can be regarded as the result of "summing up" the action (3.2) over all possible "hidden" trajectories[7]. In the action (4.2) tree parameters ($l_1 l_2 l_3 = \mathbf{l}$) mark some possible trajectory $L(\mathbf{l})$, and $P(\mathbf{l})$ is some "weight"- function, which characterize the distribution over hidden trajectories in given system and guarantee the invariance of action (4.2) not only by the coordinate transformations ($X$-transformations), but also by the "$\mathbf{l}$-transformation" $\mathbf{l} \to \mathbf{l}'$, which corresponds to changing the method of "hidden" trajectories parameterization. One can say, that the "$\mathbf{l}$-invariant": $P(\mathbf{l}) d^3\mathbf{l}$ plays in the action (4.2) roughly the same role, as the "$X$-invariant" $\sqrt{-g}\, d^4x$ in the action $\int R\sqrt{-g}\, d^4x$ for the gravitational field in the GR.

Let's imagine for instance two *different* gravitating systems A and B in GR, which are described in *the same* coordinates $X$. In this case we have

$$(\sqrt{-g})_A (x) \neq (\sqrt{-g})_B (x)$$

as per GR. Let's now imagine two *different* quantum systems A and B, described with the using of *the same* method of $\mathbf{l}$ - parameterization. We'll assume that in this case should be

$$P_A(\mathbf{l}) \neq P_B(\mathbf{l})$$

i.e. we'll suppose that $P(\mathbf{l})$ is *the physical characteristic* of systems and not only the auxiliary mathematical value, introduced in order to guarantee the $\mathbf{l}$-invariance of (4.2)[8].

Using the Dirac's $\mathbf{d}$-function, we can rewrite the (4.2) in more convenient form

$$I = \int d^3\mathbf{l}\, P(\mathbf{l}) \int_{L(\mathbf{l})} ds \int d^4x\, \mathbf{d}_4(x^i - z^i(\mathbf{l},s))\{\mathbf{m}(x) + eA_i(x)u^i(x)\} \tag{4.2'}$$

where $z^i(\mathbf{l}, s)$ is the point on the world line $L(\mathbf{l})$ in the moment $s$ of proper time.

In this article we'll regard only the simplest case when the motion of particles, as in the classical mechanics, proceeds from the past to the future, or, in other words, that the time $t$ increases monotonously along all world lines. In order the similar simplification to take place, the quantum mass must be rigorous positive that is possible only for relatively weak quantum effects, when the quantum term $(\mathbf{a}R)^{1/2}$ is small as compared to the classical mass $m\mathbf{b}$. In this case there is one-to-one conformity $(s, l^1, l^2, l^3) \Leftrightarrow (z^0, z^1, z^2, z^3)$ and we can pass from $\int ds\, d^3\mathbf{l}$ to the integral over the four-dimensional space. This gives

---

[7] Really, one can satisfy oneself, that
1) the action (4.2) could be transformed in the integral over 4-dimensional space-time,
2) for the action (4.2) there exist no "fry-field-equation" and
3) no from the possible "hidden" trajectories in (4.2) is not outstanding.

[8] One can show, that this assumption is not only possible, but also necessary, since in opposite case one should have the same distributions of probabilities for the different systems (at least in the non-relativistic limit).

$$I = \int d^4x \sqrt{-g(z)} P(z) \frac{\partial(s,\mathbf{l})}{\partial(z^0,\bar{z})} \frac{1}{\sqrt{-g(z)}} \int d^4x \mathbf{d}_4(x^i - z^i(\mathbf{l},s))\{\mathbf{m}(x) + eA_i(x)u^i(x)\} =$$
$$\int d^4x \sqrt{-g(z)} \mathbf{n}(z) \int d^4x \mathbf{d}_4(x^i - z^i(\mathbf{l},s))\{\mathbf{m}(x) + eA_i(x)u^i(x)\} \tag{4.3}$$

where

$$\mathbf{n}(z) = \frac{\partial(s,\mathbf{l})}{\partial(z^0,\bar{z})} \frac{1}{\sqrt{-g(z)}} P(z) \tag{4.4}$$

Varying the (4.3) with respect to the $k_i$ and taking into account the expression (A.7) for inner curvature $R$ we have

$$\mathbf{d}_k I = \int d^4x \sqrt{-g(z)} \mathbf{n}(z) \int d^4x \mathbf{d}_4(x^i - z^i(\mathbf{l},s)) \frac{-3\mathbf{a}}{\mathbf{m}} [(-\mathbf{d}k^i)_{;i} + 2k^i \mathbf{d}k_i] =$$
$$\int d^4x \sqrt{-g(z)} \frac{\mathbf{n}(z)}{\mathbf{m}(z)} (-3\mathbf{a}) [(-\mathbf{d}k^i)_{;i} + 2k^i \mathbf{d}k_i] \tag{4.5}$$

The expression (4.5) leads to the simple equations for the $k$-field.
$$(\mathbf{n}/\mathbf{m})_{,i} + 2(\mathbf{n}/\mathbf{m})k_i = 0 \quad \text{or} \quad \mathbf{r}_{,i} + 2\mathbf{r}k_i = 0 \tag{4.6}$$
where $\mathbf{r} = (\mathbf{n}/\mathbf{m})$ is the scalar function, which has the power $-2$ under the scale transformations. Furthermore, we'll call this function "4-density", meaning its invariability under 4-dimensional coordinate transformations. Taking into account the determination of co-covariant derivative (A 12) we can rewrite (4.6) also in the co-covariant form
$$\mathbf{r}_{*i} = 0 \tag{4.7}$$

From the equation (4.6) the impotent conclusion follows:

Since the value $\mathbf{r}_{,i}/\mathbf{r}$ is a gradient of scalar, then the scale vector $k_{,i}$ of inner space must be a gradient of scalar as well. In other words, the inner space as well as the external one must be integrable – this fact was earlier only implied and now emerged as a conclusion of equations for $k$-field. Thus in the inner space as well as in external one we can enter instead of scale vector $k_i$ a scale factor $b$, which contains (alongside with $g_{ik}$) the whole information about properties of IW-space.

Let's now consider the ensemble of $N$ identically prepared systems, which can be distinguished only by the "hidden" trajectories of test particles. In order to connect the 2-density $\mathbf{r}$ with the probability's density, let's postulate the

**Hypothesis 1.** *The integration $\int d^3\mathbf{l}$ in (4.2) is equivalent to the summation over the "hidden" trajectories in the ensemble of identical systems.*

In other word, we assume, that by enough large $N$ the distribution of hidden trajectories in ensemble is a "good discrete approximation" of continual distribution in (4.2). Let's

$$I_{[N]} = \sum_{k=1}^{N} \{-\int_{L_K} \mathbf{m} ds - e \int_{L_K} A_i dx^i\} \tag{4.8}$$

is an action of ensemble, consisting of $N$ identical systems ($L_k$ is here the "hidden" trajectory in the system with number $k$). In accordance with our assumption in the limit $N \to +\infty$ we'll have

$$I_\infty = \lim_{N \to +\infty} \frac{I_{[N]}}{N} \propto I \tag{4.9}$$

where $I$ is the action (4.2).

Variation of $I_\infty$ with respect to the k-field and passage from the sum over the particles to an integral over the 3-dimensional space give

$$dI_\infty = \lim_{N\to+\infty} \frac{1}{N} \int d^4z \sqrt{-g} \frac{\boldsymbol{n}_{[N]}}{E\sqrt{g_{00}}} (-3\boldsymbol{a})[(-\boldsymbol{d}k^i)_{;i} + 2k^i \boldsymbol{d}k_i] \qquad (4.10)$$

where the value $\boldsymbol{n}_{[N]}$ is a number density of particles in the ensemble of $N$ identical systems, which can be defined from the relation

$$dn_{[N]} = \boldsymbol{n}_{[N]} \sqrt{g} dV \qquad (4.11)$$

for a number of hidden world lines, passing through the element of volume $\sqrt{g}dV$ [9].

Let in some moment $t$ in every systems of ensemble one performs the measurement of coordinates. Since we assume that such measurement gives us the real location of particle, then the number $dn_{[N]}$ from (4.11) is simply the number of test particles, which were found in the element $\sqrt{g}dV$ and the ratio $dn_{[N]}/N$ is consistently the probability, that the particle from some choused at random system will be found in this element of volume. From here it is obvious that one should regard the value

$$\boldsymbol{w} = \lim_{N\to+\infty} \frac{\boldsymbol{n}_{[N]}}{N} \qquad (4.12)$$

as the density of probability in the ensemble of infinite number of identical systems.

Putting the variation $dI_\infty$ in (4.10) equal to zero, we come to the following equations

$$(\boldsymbol{w}/E\sqrt{g_{00}})_{,i} + 2(\boldsymbol{w}/E\sqrt{g_{00}})k_i = 0 \qquad (4.13)$$

that together with (4.6) gives

$$\boldsymbol{r} \propto \frac{\boldsymbol{w}}{E\sqrt{g_{00}}} = -\frac{\boldsymbol{w}}{(S_0 + eA_0)\sqrt{g_{00}}}. \qquad (4.14)$$

From the equations (4.6) and from the definition (A.16) of scale factor $b$ we have

$$|\boldsymbol{r}| \propto b^2 \qquad (4.15)$$

i.e. the result connecting the main geometric characteristic of inner space with 4-density $\boldsymbol{r}$ and – by means of (4.14) – with density of probability $\boldsymbol{w}$ in the ensemble of identical systems. In particular, in atomic system of units ($\boldsymbol{b} \equiv 1$), for the flat space-time ($g_{00} \approx 1$), non-relativistic speeds ($u^0 \approx 1$) and weak quantum effects $\boldsymbol{m} \approx m\boldsymbol{b}$ from (4.15) and (4.14) we can obtain

$$\boldsymbol{w} \propto b^2 \qquad (4.16)$$

or in other words, the density of probability is proportional to the square of scale factor.

Finally, let's note, that the equations of motion (3.6) can be deduced not only from the action (3.2), but also from the action (4.2) by means of varying with respect to $\boldsymbol{d}x^i(\boldsymbol{l})$. Thus, one can say, that the action (3.2) is superfluous and that all of data about the motion of test particle and about the dynamics of "k-field" in the action (4.2) is contained, which as the "real" action of quantum particle can be considered.

## 5. The conservation laws.

The equations (4.6) don't solve the task of motion of quantum particle. They don't give us the scale vector $k_i$ in the transparent form, but only connect it with 4-density $\boldsymbol{r}$, which itself needs a definition. As it will be shown in further, such additional equation can be obtained from the conservation laws for the action (4.2). There must be three of such laws, which correspond to invariance of action (4.2) firstly, under transformations of curvilinear coordinates, secondly, under gauge transformations for potentials $A_i$ and thirdly, under the scale transformations.

---

[9] here $g$ is the determinant of metric tensor of 3-dimensional space

*5.1. The equation of continuity.*

We well proceed from the action (4.2) and consider first local infinitesimal transformations of coordinates

$$x^i \to x^i + \boldsymbol{x}^i \tag{5.1.1}$$

For variations of main values under transformations (5.1) we have

$$d g_{ik} = -\boldsymbol{x}_{i;k} - \boldsymbol{x}_{k;i}, \; d\boldsymbol{b} = -\boldsymbol{b}_{,i}\boldsymbol{x}^i, \; dA_i = -A_j \boldsymbol{x}^j_{;i} - A_{i;j}\boldsymbol{x}^j, dk_i = -k_j \boldsymbol{x}^j_{;i} - k_{i;j}\boldsymbol{x}^j \tag{5.1.2}$$

In order to obtain the conservation law, corresponding to transformations (5.1.1) it is necessary to substitute (5.1.2) in the expression

$$\boldsymbol{d}_g I + \boldsymbol{d}_k I + \boldsymbol{d}_b I + \boldsymbol{d}_A I = 0 \tag{5.1.3}$$

where $\boldsymbol{d}_g I, \boldsymbol{d}_k I, \boldsymbol{d}_b I, \boldsymbol{d}_A I$ are variations of action (4.2) in all field quantities. A calculation of variations, appearing in (5.1.3), gives

$$\boldsymbol{d}_g I = -\frac{1}{2}\int d^4 x \sqrt{-g}\, \boldsymbol{r}(-p_i p_k + \boldsymbol{a} R_{ik}) dg^{ik},$$

$$\boldsymbol{d}_k I = -3\boldsymbol{a}\int d^4 x \sqrt{-g}\,(\boldsymbol{r}_{,i} + 2\boldsymbol{r} k_i) dk^i, \tag{5.1.4}$$

$$\boldsymbol{d}_b I = -\int d^4 x \sqrt{-g}\, \boldsymbol{r} m^2 \boldsymbol{b} d\boldsymbol{b},$$

$$\boldsymbol{d}_A I = -\int d^4 x \sqrt{-g}\, \boldsymbol{r} e p^i dA_i$$

where $p_i = \boldsymbol{m} u_i$ is an impulse of particle, and $R_{ik}$ is the Ricci tensor (see (A5)) of inner space. Substituting (5.1.2) into (5.1.3) we obtain after some transformations

$$[\boldsymbol{r}(-p^i p^k + \boldsymbol{a} R^{ik})]_{;k} + \boldsymbol{r} m^2 \boldsymbol{b}\boldsymbol{b}_{,k} g^{ik} - e(\boldsymbol{r} p^k A^i)_{;k} + e\boldsymbol{r} p_k A^{k;i} = 0 \tag{5.1.5}$$

or with taking into account of (4.14) in co-covariant form

$$(-p^i p^k + \boldsymbol{a} R^{ik})_{*k} + m^2 \boldsymbol{b}\boldsymbol{b}_{*k} g^{ik} - e p^k_{*k} A^i + e p^k F^i{}_k = 0 \tag{5.1.6}$$

In order to show (5.1.6) in a more transparent form, we put in (5.1.6) $p_i = -(S_{,i} + eA_i)$, take into account the relation $g^{ik}(S_{,i} + eA_i)(S_{,k} + eA_k)_{*m} = 2m^2 \boldsymbol{b}\boldsymbol{b}_{*m} + \boldsymbol{a} R_{*m}$, following from the Hamilton-Jacoby's equation (3.8) and use the generalized Bianci identities (A.20). As result we have

$$S^{;i}(S^{;k} + eA^k)_{*k} = 0 \Leftrightarrow (S^{;k} + eA^k)_{*k} = 0 \Leftrightarrow p^k{}_{*k} = 0 \tag{5.1.7}$$

that we'll call "the quantum equation of continuity".

*5.2. The gauge invariance.*

Let us now make a small gauge transformation of potentials $A_i$

$$A_i \to A_i + f_{,i} \tag{5.2.1}$$

where a function $f$ is infinitesimal. Under such transformation all other field values remain unchanged. Substituting $dA_i = f_{,i}, dg_{ik} = 0, dk_{,i} = 0, d\boldsymbol{b} = 0$ in (5.1.3) we obtain at once "the law of charge conservation"

$$J^i{}_{;i} = J^i{}_{*i} = 0 \tag{5.2.2}$$

where

$$J^i = e\bm{r}p^i = e\frac{\bm{w}}{\sqrt{g_{00}}}\frac{dx^i}{dx^o} \qquad (5.2.3)$$

can be regarded as a charge-current vector in ensemble of identical systems. It is not hard to see, that (5.2.2) is exactly equivalent to (5.1.7). For this it is enough to use in (5.2.2) the determination (5.2.3) and to take into account the field equations (4.6) as well. So the coordinate invariance and the gauge one lead to the same conservation law or, more precisely, the corresponding laws are equivalent because of the equations (4.6) for the scale vector $k_i$.

*5.3. The scale invariance.*
Under small transformations of scale we have
$$\bm{d}g_{ik} = -2(1+\bm{e})g_{ik}, \bm{d}b = -(1+\bm{e})\bm{b}, \bm{d}k_i = \bm{e}_{,i}, \bm{d}A_i = 0 \qquad (5.3.1)$$
where $\bm{e}(x)$ is an infinitesimal coordinate function. Substituting of these variations into (5.1.3) gives the following conservation law
$$g^{ik}\bm{r}_{*i*k} = 0 \qquad (5.3.2)$$
which however as well as (5.2.2) does not carry anything new, since it is an obvious consequence of (4.6).

Thus the three kinds of invariance give us only one supplementary equation (namely the equation (5.1.7)), which can be used to complete the system of equations.

## 6. The complete system of equation.

As the basic equation, describing a motion of quantum particle, the Hamilton-Jacoby's equation (3.8) can be regarded
$$g^{lm}(S_{,l} + eA_l)(S_{,m} + eA_m) = \bm{m}^2 \qquad (6.1)$$
that is equivalent to the equations of motion (3.6). The vector $k_i$, which is present in the right part of (6.1), can be found from the field equations (4.6) or from the equivalent equation (4.15) for the scale factor $b$
$$|\bm{r}| \propto b^2 \qquad (6.2)$$
The scalar $\bm{r}$ can be defined in its turn from the conservation law (5.1.6), which in combination with (4.6) and (3.8) gives a complete system for $k_i$, $\bm{r}$ and $S$. In fact however it is more convenient instead of (5.1.6) and (6.2) to use (5.1.7)
$$(S^{,k} + eA^k)_{*k} = 0 \qquad (6.3)$$
which is a consequence of (5.1.6) and (4.6) and doesn't contain $\bm{r}$ anymore. Thus the pair of equation (6.1) and (6.3) allows to find the action $S$ and the scale factor $b$ and the equation (6.2) allows to connect $b$ with the 4-density $\bm{r}$. Furthermore we'll need also the equation (4.14)
$$\bm{r} \propto \frac{\bm{w}}{E\sqrt{g_{00}}} = -\frac{\bm{w}}{(S_0 + eA_0)\sqrt{g_{00}}} \qquad (6.4)$$
in order to connect the 4-density $\bm{r}$ (and hence the scale factor $b$) with the probability's density of ensemble.

# 7. Hidden variables and the quantum mechanics.

In order to establish a connection between the quantum mechanics and our model, let's regard an ensemble of identical systems with same $S$ and $b$, which is equivalent to some quantum-mechanical ensemble. The 4-density $r$, the density of probability $w$ and the "charge-current vector" $J_k$ of ensemble can be found from (4.14), (4.15) and (5.2.3), and the evolution of $S$ and $b$-values (with help of which $r$, $w$ and $J_k$ are expressed) is described by means of equations (3.8) and (5.1.7), which in atomic system of units take the form

$$g^{lm}(S_{,l} + eA_l)(S_{,m} + eA_m) = m^2 + 6a\left[\frac{b_{;l}}{b}\frac{b^{;l}}{b} + \left(\frac{b^{;l}}{b}\right)_{;l}\right], \tag{7.1.a}$$

$$(S^{;l} + eA^l)_{;l} + 2\frac{b_{;l}}{b}(S^{;l} + eA^l) = 0 \tag{7.1.b}$$

Let us now construct an "artificial" complex function $\Psi$, which is related with $S$ and $b$ by

$$\Psi = b\exp(iS/\sqrt{6a}) \tag{7.2}$$

and besides to put

$$a = \hbar^2/6, \tag{7.3}$$

since $a$ is the free parameter of model and can be chosen in arbitrary way. It is easy to show, that in this case the 4-density $r$ and the charge-current vector $J^k$ are related with $\Psi$ in the following way

$$r \propto \frac{w}{mu_0\sqrt{g_{00}}} \propto \Psi\Psi^* \tag{7.4.a}$$

$$J^k \propto i[\Psi^*(D^k + \frac{ie}{\hbar}A^k)\Psi - \Psi(D^k + \frac{ie}{\hbar}A^k)\Psi^*] \tag{7.4.b}$$

and that the pair of equations (6.1) can be changed by the equivalent complex equation

$$(i\hbar D^k - eA^k)(i\hbar D_k - eA_k)\Psi = m^2\Psi, \tag{7.5}$$

where $D_k$ is a covariant derivative. So, instead of describing the ensemble in terms of pair of real values $S$ and $b$ and pair of real equations (7.1), we can use one complex function $\Psi$ and one complex equation (7.5), expressing the values $r$ and $J^k$ accordingly to the (7.4). From these it is obvious, that the "artificial" function $\Psi$ has all properties of wave functions $\Psi_{qm}$ of quantum mechanics. In particular, the equation (7.4.a) is simply a relativistic analog of Born's interpretation of wave function; the equation (7.4.b) is the quantum-mechanical expression for the charge-current vector in a quantum ensemble and (7.5) is the basic equation of quantum mechanics i.e. the relativistic wave equation. Thus we can conclude, that the function $\Psi$, which is connected with $S$ and $b$ accordingly the (7.2), is identical to the quantum-mechanical wave function of ensemble and that all predictions for this ensemble, which are deduced in the basis of hidden variables' conception must coincide with the predictions of quantum mechanics.

Let us finally write the quantum-mechanical functions $\Psi_{qm}$ in the form

$$\Psi_{qm} = r\exp(is) \tag{7.6}$$

where $s$ is the phase and $r$ is the module of wave function. The following conclusions are obvious:

1. The action $S$, which describes a motion of test particle, coincides with accuracy to additive constant with the phase of quantum-mechanical wave function multiplied on the Plank's constant $\hbar$ [10].
$$S = s\hbar + const \qquad (7.7)$$
2. The scale factor $b$ of inner space coincides with the module $r$ of wave function with accuracy to a constant multiplier.
$$b = const \cdot r \qquad (7.8)$$
3. The Hamilton-Jacoby's equation (7.1.a) coincides with the real part of wave equation of quantum mechanics and the equation of continuity (7.1.b) with the imaginary one.

From here we have the simple method of determination for the values $S$ and $b$ - and hence for the values $k_i$ and $u_i$ - about which it was been said in the section 4. In order to determine the action $S$ and the scale factor $b$ of some individual system "$a$" it is enough simply to determine the quantum-mechanical wave function $\Psi_{qm}$ of ensemble "$A$" to which the system "$a$" belongs and than to connect $\Psi_{qm}$ with $S$ and $b$ by means of (7.7) and (7.8). It is obvious, that in this case there is no necessity to make the measurement over "$a$" – since instead of "$a$" the other systems of "$A$" can be used – and hence the station of "$a$" by this method of "measuring" of $S$ and $b$ can remain undisturbed.

Thus the task of reduction of quantum ensembles dynamics to dynamics of individual systems is solved. One can notice a large progress as compared to the old Bohm's theory. Bohm started with the wave equation, we finished with it. For Bohm the wave equation was a needed postulate, without which he could not co-coordinate his theory with the quantum mechanics; for us the wave equation is a natural conclusion from our basic foundations. Finally, the statements 1-3, connecting the basic values and expressions of quantum mechanics and these from the THV, appeared in our article as natural final results unlike the Bohm's theory [2], where they played a role of very mysterious postulates.

## Acknowledgments.

The author is grateful to Doctor Roderich Tumulka (Muenchener Institut fuer Mathematik) for fruitful discussion.

## Appendix A.

In the basis of Weyl's geometry lie the following axioms:
1. The change of vector $x^i$ by parallel displacement can be defined as
$$dx^i = -\Gamma^i{}_{kl} x^k dx^l$$
(A.1)
where $\Gamma^i{}_{kl} = \Gamma^i{}_{lk}$ the symmetric affine connection.
2. The change in length of vector by parallel transport is given by
$$d(x^i x_i) = 2(x^i x_i) k_l dx^l$$
(A.2)
where $g_{ik}$ is the metric tensor and $k_l$ the scale vector of Weyl space. The values $g_{ik}$ and $k_l$ are basic objects of Weyl's geometry and the rest of values of Weyl space can be expressed by means of $g_{ik}$ and $k_l$. In particular, for $\Gamma^i{}_{kl}$ it can be easily shown.

---

[10] From the (7.7) one can derive at once the relativistic "guiding equation" i.e.
$$mu^i = eA^i + \hbar \, \text{Im}\{\Psi^i \Psi^* / \Psi \Psi^*\}$$

$$\Gamma^i{}_{kl} = \boldsymbol{g}^i{}_{kl} - (\boldsymbol{d}^i_k k_l + \boldsymbol{d}^i_l k_k - g_{kl} k^i)$$
(A.3)

where $\boldsymbol{g}^i{}_{kl}$ are the Christoffel symbols, defined in terms of $g_{ik}$ as in Riemannian geometry.

A generalized curvature tensor in Weyl space can be written as follows.

$$R^i{}_{klm} = \Gamma^i{}_{kl,m} - \Gamma^i{}_{km,l} + \Gamma^n{}_{kl}\Gamma^i{}_{nm} - \Gamma^n{}_{km}\Gamma^i{}_{nl}$$
(A.4)

that is quite analogous to the Riemannian spaces case. The curvature tensor has three nontrivial contractions (from which only two are independent)

$$R^{(1)}_{mn} = R^k{}_{mkn} = r_{mn} - 2(k_{m;n} - k_{n;m}) - (k_{m;n} + k_{n;m}) - g_{mn} k^l{}_{;l} - 2k_m k_n + 2g_{mn} k^l k_l,$$

$$R^{(2)}_{mn} = R^{.l}_{m.nl} = R^{(1)}_{mn} - 2F_{mn},$$
(A.5)

$$R^{(3)}_{mn} = R^l{}_{lmn} = 4F_{mn}$$

where

$$F_{mn} = k_{n;m} - k_{m;n}$$
(A.6)

and $r_{mn}$ is defined in terms of $g_{mn}$ as the Ricci tensor of Riemannian geometry. The symbol ";" marcs here and hereafter the ordinary covariant derivative defined by means of Christoffel symbols $\boldsymbol{g}^i{}_{kl}$.

The scalar curvature of Weyl space is the only one

$$R = g^{ik} R^{(1)}_{ik} = g^{ik} R^{(2)}_{ik} = r - 6k^l{}_{;l} + 6k^l k_l$$
(A.7)

where $r = g^{ik} r_{ik}$ is defined by means of $g_{ik}$ as a scalar curvature of Riemannian space.

Under a general scale transformation

$$ds \to d\bar{s} = l(x) ds$$

(A.8) the metric tensor $g_{ik}$ must change accordingly to the law.

$$g_{ik} \to \bar{g}_{ik} = l^2(x) g_{ik}$$

(A.9) since $ds^2 = g_{ik} dx^i dx^k$ and since $dx^i$ are not affected by the scale transformation.

From (A2) it can be shown, that under the transformations (A8) $k_i$ changes as

$$k_i \to \bar{k}_i = k_i + (\ln l)_{,i} \qquad (A.10)$$

Using (A9) and (A10) it can be proved that the affine connection $\Gamma^i{}_{kl}$ is invariant under scale transformations. In the same manner it can be easily shown that $R^i{}_{klm}$ and the contracted tensors $R^{(1)}_{ik}$, $R^{(2)}_{ik}$ and $R^{(3)}_{ik}$ are also scale invariant.

The passage from the Riemannian space to the Weyl space demands a passage from a conception of tensor to the one of co-tensor. Let $A$ is a tensor of arbitrary rank and let $A$ changes under scale transformations according to the law

$$A \to \bar{A} = l^n A \qquad (A.11)$$

Then $A$ is called the co-tensor of power $n$. One denotes the power of tensor as $\Pi(A)$ i.e. the expression

$$\Pi(A) = n$$

denotes, that the co-tensor $A$ has the power $n$. If $\Pi(A)$ is zero, it is called an in-tensor. So the values $R^i{}_{klm}$ and $R^{(i)}_{ik}$ are in-tensors, $g_{ik}$ is a co-tensor of power +2 (see (A9)), and $g^{ik}$ is a co-tensor of power –2.

It should be mentioned that not all tensors are co-tensors. For example $r^i{}_{klm}$ and $r_{ik}$ are tensors under coordinate transformations but don't transform like (A12) under scale transformations (A8).

A product of co-tensors is again a co-tensor. If $A$ and $B$ are co-tensors of power $n_1$ and $n_2$ respectively, then a co-tensor $C = AB$ is a co-tensor of power $n = n_1 + n_2$. From this it is clear that $g^{ik}$ has the power $-2$[11] as well as the scale curvature $R = g^{ik} R_{ik}$.

The extension of conception of tensor to the one of co-tensor requires a corresponding generalization of operation of differentiation, since a usual covariant derivative of tensor is not in general a co-tensor. Let $S, V, T$ are co-tensors of power $n$ and of ranks 0,1 and 2 respectively. The co-covariant derivative "*" of this values is defined as follows (the generalization for the higher rank's tensors is obvious)

$$S_{*m} = S_{,m} - n k_m S,$$
$$V^m{}_{*n} = V^m{}_{,n} + \Gamma^m{}_{nl} V^l - n k_n V^m, \qquad (A.12)$$
$$V_{m*n} = V_{m,n} - \Gamma^l{}_{mn} V_l - n k_n V_m,$$
$$T^{mn}{}_{*l} = T^{mn}{}_{,l} + \Gamma^m{}_{kl} T^{kn} + \Gamma^n{}_{kl} T^{km} - n k_l T^{mn}$$

From these definitions it is easily to show that the co-covariant derivative of co-tensor of power $n$ is again a co-tensor of power $n$ and that a product of co-tensors $A$ and $B$ of arbitrary ranks and powers satisfies the usual product law

$$(AB)_{*i} = A_{*i} B + A B_{*i} \qquad (A.13)$$

From the definition (A14) one can see that the metric tensor $g_{ik}$ satisfies the relations

$$g_{ik*l} = 0, \; g^{ik}{}_{*l} = 0 \qquad (A.14)$$

Let now a change in length under a parallel displacement is integrable i.e. doesn't depend on a path of transport. In this case the $k_m$ must satisfy the scale-covariant condition of integrability

$$k_{m;n} - k_{n;m} = 0 \qquad (A.15)$$

From (A.15) we conclude that the scale vector of such integrable Weyl space (IW-space) is a gradient of scalar function. In this case it is convenient to use "the scale factor $b$", which defines all properties of IW-space and is connected with $k_m$ as follows

$$k_m = -\frac{b_{,m}}{b} \qquad (A.16)$$

Under scale transformations (A8) the scale factor $b$ changes as

$$b \to \bar{b} = b l^{-1} \qquad (A.17)$$

i.e. $b$ is a co-scalar of power –1. From this and from (A.16) it is easy to see that the scale factor $b$ must satisfy the equation

$$b_{*i} = 0 \qquad (A.18)$$

or in other words to be co-covariant constant. It is obvious, that the scale factor $b$ is defined only with accuracy to a constant factor or in other words with accuracy to global scale transformations.

---

[11] since $g^{ik} g_{kl} = \boldsymbol{d}^i_l$ and $\Pi(\boldsymbol{d}^i_l) = 0$

In the IW-space a number of contractions of $R^i{}_{klm}$ reduces from three to one and in this case we have

$$R^{(1)}_{ik} = R^{(2)}_{ik} = R_{ik},$$
$$R^{(3)}_{ik} = 0 \qquad (A.19)$$

Let us note finally that $R_{ik}$ and $R$ satisfy the generalized Bianci identities

$$R^i_{k*i} - \frac{1}{2} R_{*k} = 0 \qquad (A.20)$$

### Appendix B.

Hear the more precise definition of "identity" of systems is given.

Let $M$ and $N$ are two non-interacting systems, every of which consist of the source of external fields $A_i$ and $g_{ik}$ and of the test particle, moving in the field of source. Let further

$$\{e^{(M)}, m\mathbf{b}^{(M)}, A_i^{(M)}, g_{ik}^{(M)}\}_{(X_{(M)}, G_{(M)}, \mathbf{b}_{(M)})} \qquad (B1)$$

is the set of values, which characterize the system $M$, written in some coordinates $X_{(M)}$, in some electromagnetic gauge $G_{(M)}$ and in some system of units $\mathbf{b}_{(M)}$. The system $N$ is called "identical to the system $M$", if there is such "method of description" $(X_{(N)}, \mathbf{b}_{(N)}, G_{(N)})$ for the system $N$, for which a character of dependence of values

$$\{e^{(N)}, m\mathbf{b}^{(N)}, A_i^{(N)}, g_{ik}^{(N)}\}_{(X_{(N)}, \mathbf{b}_{(N)}, G_{(N)})} \qquad (B2)$$

from the coordinates $X_{(N)}$ is the same as the (B1) for the $M$-system. If now to marc the coordinates $X_{(N)}$ with the same symbols, as the coordinates $X_{(M)}$ (simply to suppose $X_{(N)} = X_{(M)} = X$), then the "identity" of systems $M$ and $N$ can be written as follows:

$$e^{(N)} = e^{(M)}, \; (m\mathbf{b})^{(N)}(x) = (m\mathbf{b})^{(M)}(x),$$
$$g_{ik}^{(N)}(x) = g_{ik}^{(M)}(x), A_i^{(N)}(x) = A_i^{(M)}(x), \qquad (B3)$$

and as the consequence of "classical identity"

$$S^{(N)}(x) = S^{(M)}(x), \; k_i^{(N)}(x) = k_i^{(M)}(x)$$

(B4) i.e. the identity of quantum characteristics (see section 4).

From the external observer's point of view all identical systems are really "identical"(undistinguished), but from the point of view of theoretician, explaining the observational data, the distinction between the systems can be still introduced. In spite of the external "identity" one can admit, that the test particles from the different systems move "in reality" along the different "hidden" trajectories, which corresponding to the different initial conditions for the coordinate $X$.

### Literature.